# Geometrical design of thin film PV modules for improved shade tolerance and performance

Sourabh Dongaonkar and Muhammad A Alam

School of Electrical and Computer Engineering, Purdue University, West Lafayette, IN

**ABSTRACT**

Partial shading in photovoltaic modules is an important reliability and performance concern for all photovoltaic technologies. In this paper, we show how cell geometry can be used as a design variable for improved performance and resilience towards partial shading in monolithic thin film photovoltaic (TFPV) modules. We use circuit simulations to illustrate the geometrical aspects of partial shading in typical TFPV modules with rectangular cells, and formulate rules for shade tolerant design. We show that the problem of partial shading can be overcome by modifying the cell shape and orientation, while preserving the module shape and output characteristics. We discuss two geometrical designs with cells arranged in radial and spiral patterns, which (a) prevent the reverse breakdown of partially shaded cells, (b) improve the overall power output under partial shading, and (c) in case of spiral design, improve the module efficiency by reducing sheet resistance losses. We compare these designs quantitatively using realistic parameters, and discuss the practical approaches to their implementation.

*Keywords* – partial shading, thin film PV module, module design, module efficiency, sheet resistance.

## 1. INTRODUCTION

In the last decade, thin film photovoltaic technology has transferred from lab to commercial scale, and is now an important component of the photovoltaic market [1]. Moreover, TFPV promises to offer important advantages in the growing building integrated PV (BIPV) sector [2]. This change has been driven by the advances in the large scale manufacturing of TFPV, including the monolithic fabrication technique [3]. This approach involves successive deposition of contact and semiconductor layers on large area glass or flexible substrates, interspersed with scribing steps to form a series connected module, as shown in Figure 1 [4], [5]. The resulting module configuration has thin rectangular cells next to each other, connected in series with metal interconnects (Figure 1).This monolithically integrated manufacturing process keeps throughput high, and provides important cost advantage for TFPV technologies.

This monolithic fabrication, however, introduces a unique set of challenges for TFPV module operation under real world operating conditions. One such challenge arises in case of partial shading of the modules by shadows cast by nearby objects and structures [6], [7]. This problem is by no means limited to TFPV technologies, having been first observed for crystalline cells in space applications [8]. In case of crystalline PV, however, the manufacturing process allows for incorporation of bypass diodes inside the module [9], or alternate wiring schemes for cells [10], [11], which can mitigate the impact of shading. These approaches are not easily transferrable to monolithic TFPV modules, because the scribing based interconnection scheme makes it difficult to integrate bypass diodes [12], or alternate wiring schemes [13]. Another aspect of partial shading unique to TFPV modules is related to the rectangular geometry of individual cells. The 2D analysis reported in [14] shows that the interplay between rectangular shadows and (series-connected) thin long rectangular cells dictates the degree of reverse stress experienced by shaded cells [14], with some configurations leading to extreme reverse bias and causing permanent damage, while other shadows are harmless.

In this paper, we show that this monolithic fabrication technique can be adapted to create different cell geometries for improving shade tolerance, while preserving (even improving) module performance. These geometrical transformations, however, must ensure that (i) the series connection of cells is preserved, (ii) all cell areas must remain identical, and (iii) the rectangular module shape is unchanged. These design constraints ensure that the nominal module output remains unchanged after geometrical transformation of cell shape. We use 2D SPICE circuit simulation to demonstrate how this flexibility in choosing cell shape can not only alleviate the problem of partial shading in TFPV modules, but also improve module performance.

We begin in Section 2 by describing the simulation framework, which will be used to assess the shadow effect, as well as module performance. Next, in Section 3, we discuss the shadow effect in typical TFPV modules with rectangular cells. We highlight the key reliability issues of partial shading, and formulate the design rules for shade tolerant design. In Section 4, we discuss two geometrical shade-tolerant designs for TFPV modules, and evaluate their reliability and performance under various shading scenarios. Next, in Section 5, we analyze the sheet resistance losses in these non-rectangular cells for assessing the efficiency of these new module designs. Finally, in Section 6, we will summarize the results, and discuss the practical considerations of implementing the proposed designs.

## 2. SIMULATION FRAMEWORK

In order to study the shadow effect in realistic systems, we consider a string of series connected modules (each with an external bypass diode [15]) connected to a string inverter, as shown in Figure 2(a). The number of modules in the string is adjusted to obtain the desired string DC output voltage, which is assumed to be kept fixed at all times [16]. We will evaluate the string output power, as well as voltage across shaded cells, when one of the modules in the string is partially shaded. For the partially shaded module, we consider rectangular shadow at the bottom left corner (see Figure 2(b)), and all possible shade configurations are explored by varying the shadow dimensions ($L_{sh} \times W_{sh}$) from 0 (no shade) to the module dimensions $L_{module} \times W_{module}$ (fully shaded). It is assumed that the shaded region only receives diffused light, so that the light intensity in the shaded region is 20% of the normal sunlight intensity [17].

## 2.1 Analysis of shadow effect

We had shown earlier that the partially shaded module with 2D shadow (as in Figure 2(b)), can be simulated very accurately in SPICE, by using a 1D equivalent circuit of the module [14], shown schematically in Figure 2(c). This technique is also suitable for simulating the shadow effects for arbitrary cell shapes of shade tolerant designs, as discussed in Appendix A. This approach allows us to calculate all the individual cell voltages $V_{cell}$ for all the cells in the partially shaded module (see Figure 2(c)), and find the minimum cell voltage $V_{cell}^{min}$, for different shading scenarios. The objective of shade tolerant design is to restrict $V_{cell}^{min}$ to small values, and prevent catastrophic breakdown of shaded cells [18], [19]. From this simulation, we also calculate the string output power $P_{string}$, for various shading scenarios. The shade tolerance of the designs will be evaluated by comparing the $V_{cell}^{min}$ and $P_{string}$ values for each design under all possible shading scenarios.

In order to compare the module designs, we chose a typical a-Si:H technology as a base case [20]. For simplifying the geometrical transformations, we have assumed a square module shape, with $L_{module} = W_{module} = 128 cm$, and $N_{series} = 128$ (instead of the actual $L_{module} = $

$104 cm$, $W_{module} = 120 cm$, with $N_{series} = 104$). This slight reshaping of the module changes the values marginally, but makes no difference to the conclusions drawn. In this configuration, we need 6 modules in the string to achieve the string output of 535V DC, where all modules are operating at their maximum power points. We consider the situation when one of the modules is partially shaded, and evaluate the worst case voltage developed across cells inside the shaded module, as well as the string output power under each shading scenario.

**2.2 Sheet resistance analysis**

While the equivalent circuit approach based on SPICE is sufficient for evaluating the shadow effect, a second (complementary) simulation approach is necessary to calculate the performance loss due sheet resistance for arbitrary cell geometry. We need to consider the 3D current flow (shown for typical rectangular TFPV cell in Figure 2(d)), showing the current entering from the left side of bottom metal contact, then bulk current flow in the semiconductor, and finally collected at the right edge of top TCO contact.

Assuming that the metal sheet resistance is small (see Appendix B), we can solve the continuity equation for 2D current flow $\vec{J}_{xy}$ in top TCO contact only, with photocurrent $J_{ph}$ is injected at all points of the sheet conductor (see Figure 2(e)); so that $\vec{\nabla}_{xy} \cdot \vec{J}_{xy} = J_{ph}$. Writing $\vec{J}_{xy} = \sigma \vec{\mathcal{E}}_{xy} = \vec{\mathcal{E}}_{xy}/R_\square$, with the sheet resistance $R_\square$ of the TCO layer, and $\vec{\mathcal{E}}_{xy} = -\nabla_{xy}\phi$, in terms of potential $\phi$, we get

$$\nabla^2_{xy}\phi = -J_{ph}R_\square. \tag{1}$$

This equation can be solved numerically with appropriate boundary conditions and the solution can be used to calculate the resistive power dissipation in the TCO layer as

$$P_{dis} = \int_{A_{cell}} \vec{J}_{xy} \cdot \vec{\mathcal{E}}_{xy} dS = \int_{A_{cell}} |\vec{\mathcal{E}}_{xy}|^2 / R_\square dS; \qquad (2)$$

where $\int_{A_{cell}} dS$ denotes the surface integral over the entire area of the cell $A_{cell}$. Further details regarding the numerical solution setup can be found in appendix B. With these tools, we are now ready to examine the shadow behavior, and module efficiency for various module types. We begin by analyzing conventional rectangular module geometry.

## 3. SHADOW EFFECT AND DESIGN RULES

In this section we analyze the geometrical aspects of shadow effect in typical TFPV modules with rectangular cells, and outline the design rules for shade tolerant design.

### 3.1 Rectangular design with rectangular cells

Figure 3 shows the result of the SPICE simulation analyzing the effect of partial shading, by plotting the worst case reverse stress $V_{cell}^{min}$ at the cell level, and the string power output $P_{string}$, for all possible shadow sizes. Each point on the color plot represents the effect of a shadow of certain size, and the color denotes the corresponding $V_{cell}^{min}$ (Figure 3(a)), or $P_{string}$ (Figure 3(b), for that shadow size. The simulation shows that while the worst case reverse stress occurs for small wide shadows (e.g. along the bottom edge of the module), but the external bypass diode turns on for only a fraction of shading scenarios with large shadows (see Figure 3(a),(b)), and does not prevent the worst case reverse stress [14]. Some interesting insights about the shadow effect are also apparent from the plots. First, note that a symmetric edge shadow (with all cells shaded equally, as shown in Figure 3(a)), causes no reverse stress, and relatively small loss of output power. On the other hand, an asymmetric shade at the edge (Figure 3(a)) causes reverse breakdown of shaded cells, and reduces output power dramatically. This is because, in the asymmetric case, the fully illuminated cells continue to drive the current through the shaded cells, and push them in reverse bias. Note, however, that as more cells are shaded in the asymmetric case, the stress on individual ones is reduced ($V_{cell}^{min}$ becomes less negative), because the reverse voltage is equally divided among the shaded cells.

**3.2 Geometrical design rules for shade tolerance**

Based on these observations, we can formulate a set of design rules for a shade tolerant design of a TFPV module. These can be summarized as:

1. The strong difference in effect of symmetric vs. asymmetric shading suggests that a shade tolerant design must be free from this orientation dependence.
2. The worst case of thin asymmetric shadow must be avoided to prevent permanent damage.
3. Shading of multiple cells together distributes the reverse bias, and if it can be utilized by the new design, permanent damage to shaded cells can be averted.

It is easy to see that if the rectangular cells of TFPV modules could be arranged radially (like the blades of a fan), the worst case shadow stress can be reduced, because a rectangular shadow will now cover small areas of multiple cells. This, however, cannot preserve the series connection or the rectangular module shape. Fortunately, the monolithic fabrication allows us to change the cell shape in a way, which will satisfy all these constraints, and preserve the module shape and output characteristics at the same time.

**4. SHADE TOLERANT DESIGN**

The fundamental insight behind this is the observation that a good overlap of a rectangular shadow with rectangular cell is the cause of worst case shadow stress; and, while shadows are generally rectangular (buildings, poles etc.), the cells need not be. Their shape can be modified in a way, which reduces the probability of perfect overlap between a rectangular shadow and a non-rectangular cell.

**4.1 Radial design with triangular cells**

The simplest geometry which satisfies the design constraints outlined in the previous section is formed by modifying each rectangular cell into two triangular half-cells and arranging them in a radial pattern, as shown in Figure 4(a). The current flow patterns in two types of half cells are also shown for comparison. The terminals need to be put in diagonally, as shown in the schematic and the current flows in a curved path dictated by the series connection. Note that the triangle dimensions and orientations can be chosen to ensure that all cells are of equal area, number of series connected cells is the same, and the square module dimensions are preserved (see Appendix C for details on how to generate the radial geometry).

The new design can now be assessed using the same SPICE simulation framework (see Appendix A for details). The results of the simulation are shown in Figure 4(b), which shows the color plot of $V_{cell}^{min}$ under partial shading, for all different shadow sizes for the radial design. We find that the $V_{cell}^{min}$ value for the radial design is always above $-4.7V$, thus preventing any permanent damage from partial shading [18]. Also, note that the radial symmetry ensures that there is no difference between symmetric and asymmetric shading scenarios (evident from the schematic in Figure 4(a)). The corresponding $P_{string}$ values for different shading scenarios for this design are shown in Figure 4(c), showing marked improvement for smaller shadows, and values are comparable for large shadows. Moreover, Figures 4(b),(c) show that for this radial design, only large shadows cause any significant output power loss or reverse stress, but the more likely smaller shadows are rendered practically harmless. Thus, we see that the radial design can significantly improve the shade tolerance of the module.

This shade tolerance of radial design, however, is achieved at a cost of higher resistive losses in the triangular cells. This is because the cells have to be wider near the base of the triangle, thereby increasing the path length for the photocurrent. A detailed analysis and comparison of the resistive losses is presented in Section 5. Another likely limitation of this design is the problem of oblique shadows. It is readily apparent from the schematic in Figure 4(a) that a diagonal shadow (from bottom left to top right) on this radial module will shade the two diagonal cells fully, similar to the asymmetric worst-case shading in rectangular modules. While this is an unlikely scenario for utility scale installation, where edge shadows from nearby modules are more likely [21]; it can be a concern for BIPV systems where diagonal shadows from nearby objects are possible [2].

In order to reduce the resistive losses, we note that the average cell width should be reduced, so that photocurrent flows over shorter distances before being collected. This must be done while keeping the cell area constant; which will be possible by making the cell shape longer and thinner. And, the problem posed by diagonal shadows can be avoided if the cells themselves are non-rectilinear, so that the asymmetric shading will never arise. We show next that both these objectives are achieved by a spiral arrangement of curvilinear cells.

**4.2 Spiral design with curvilinear cells**

Figure 5(a) shows the schematic of the spiral design, with the same $N_{series}$ series connected cells, with the curved positive and negative terminals. Each cell is a concave polygon with varying curvature, constructed so that their areas are identical, while preserving module shape and size. Details regarding constructing this geometry are given in the supplementary materials. These curved cells can be considered to be stretched and twisted forms of the triangles used in the radial design, arranged within the same rectangular module. Therefore, the current flow direction is also similar, as shown in Figure 5(a).

From the schematic in Figure 5(a) it is obvious that this spiral design retains the advantages of the radial design in terms of shade tolerance. This is validated from the circuit simulation result in Figure 5(b); which shows that the worst case $V_{cell}^{min}$ from shading in this spiral design is limited to $-4.2V$, and the overall number of cases which result in reverse bias is also reduced. Figure 5(c) also shows the improvement in string output power for the various shading scenarios; which also shows that external bypass diode is activated only for very large shadows. Therefore, with the spiral design it may be possible to avoid the external bypass diode altogether, enabling a truly monolithic TFPV module, and eliminate the considerable reliability issues associated with the external bypass diodes [22]. Moreover, from the schematic in Figure 5(a), it is apparent that due to the curved cell shape, the asymmetric shading problem cannot arise for rectilinear shadows of any orientation. Finally, we will show in Section 5 that the curvilinear cell shape also reduces overall resistive power loss compared to rectangular cells and improves module efficiency.

## 5. EFFICIENCY COMPARISON

In this section, we evaluate how a change in cell shape affects the normal (no shade) operating performance of the module. We demonstrate that it is possible to achieve shade tolerance without a tradeoff in nameplate module efficiency, because of reduced sheet resistance losses in non-rectangular cells of the shade tolerant design. In order to compare sheet resistance losses in cells with different geometries, we solve Equation (1) for all three geometries, and calculate the power dissipation per unit area in the sheet conductors. Color plots at the top of Figure 6 show the power dissipation per unit area in the TCO layer, obtained from the numerical solution to

Equation (1), assuming $R_{\square}^{TCO} = 10\ \Omega/sq.$, and $J_{ph}^{mpp} = 13\ mA/cm^2$. The plots show the simulation results for the sub-module schematics shown in Figures 3(a), 4(a), and 5(a), respectively.

For the rectangular geometry, the power dissipation profile is identical for all cells. The dissipation per unit area increases monotonically towards the current collecting (top) edge of each cell. This is because the regions near the current collecting edge carry the total current generated in the area below, resulting in higher power loss per unit area (see Appendix B for details). Using equation (2) for the rectangular geometry, we can get the total resistive power dissipation $P_{dis}^{rec} = 72.1\ mW$ per cell; which means for $N_{series} = 128$, $9.22\ W$ is dissipated in total in the TCO sheet resistance. For the two shade tolerant designs, the $P_{dis}$ values will be different for each cell, because shape and orientation (and hence the current flow pattern) is different for each cell. As shown in the color plots in Figure 6, the power dissipation per unit area in wider regions is higher than the thinner regions, as the current collecting edge near the wider areas collects more photocurrent. As a consequence, the wider triangles in the radial design (close to horizontal and vertical axes) dissipate twice the power compared to rectangular cells, while the dissipation in thin diagonal cells is almost equal to the rectangular case (compare the $P_{dis}^{rad}$ values with $P_{dis}^{rec}$ from the plot in Figure 6). The increased power dissipation per unit area in each cell translates into total power dissipation of $14.57W$ in the whole module with 128 cells, which is ~50% higher than the rectangular case.

This geometry dependence in resistive power dissipation is also visible in the curvilinear cells of the spiral design; where the thinner regions near the center dissipate less power compared to the wider areas towards the edges. Moreover, the pattern in $P_{dis}^{spi}$ values across different cells is also the same, and the longer diagonal cells dissipate less power compared to wider cells near the middle. Overall, however, the cells in spiral design dissipate *less* power compared even to the rectangular cells (see the $P_{dis}^{spi}$ values in the plot in Figure 6). Correspondingly, the total resistive loss in the spiral module of 128 cells is only $6.42W$, which is about 30% less than even the rectangular case. This reduction in $P_{dis}^{spi}$ stems from the fact that the perimeter of these curvilinear polygonal cells is larger than the rectangular cells, while $A_{cell}$ is the same. Therefore, the width of each cell is smaller (on an average) compared to rectangular cells, which reduces the overall power dissipation. A more detailed discussion about numerical calculation of the power dissipation in sheet resistors, and its relation to previous approaches using circuit simulations, is provided in Appendix B.

## 6. DISCUSSION AND CONCLUSIONS

In the previous sections, we have demonstrated that a geometrical approach to module design can not only alleviate the problem of partial shading in TFPV modules, but can also enhance the overall module performance. We believe this approach towards module design is practically viable, and offers attractive improvements without requiring significant tradeoffs. We would like to emphasize that while the calculations presented here were done for a typical a-Si:H solar cell, the conclusions are equally valid for all monolithic TFPV technologies. For other TFPV technologies, the exact cell characteristics including the dark and light IV behavior and reverse breakdown voltages would change, but the general conclusions about the shade tolerance of the new designs would not be affected.

From a practical standpoint, laser scribing is the most suitable technique for manufacturing of these non-conventional cell geometries. Laser scribing has been used extensively for thin film Si technologies [23], [24], and it is being actively developed for other polycrystalline TFPV technologies [5], [25], [26]. We think it should be possible to adapt them for the proposed non-rectangular cell geometries, possibly through a combined motion of the laser head and rotation of the platform carrying the module to create the non-rectangular shapes.

A potential cause of concern is that the radial and spiral designs require longer scribe lines, which may result in increased edge-shunting. It has been shown that for optimized laser scribing process, the edge shunts are not a major concern [27], and the random shunt formation across the cell surface is the dominant shunting mechanism [28]. A second concern is related to increased 'dead area' due to longer scribe lines in the shade tolerant design. Assuming typical dead region width for a laser scribes to be $300 \mu m$, we can calculate the total length of scribe lines for each geometry. We find that the percentage of dead area increases from 3% of module area, for the rectangular design, to 3.44% for the radial case, and 4.66% for the spiral case. While these increases in dead area are not insignificant, they are far outweighed by the reduced resistive loss and improved reliability of the design discussed above. Furthermore, recent techniques like point-wise interconnection [29], are imminently suitable for the shade tolerant designs, and would help in reducing the dead area further.

Finally, we note that geometrical design in different guises has been used for improving the PV performance on different levels. The prominent examples of this include various light trapping schemes (at the cell level) [30], and the recent 3DPV approach to module arrangement (at the module level) [31]. We feel that the shade tolerant design proposed in this work is in the same vein, and utilizes geometry in a unique way to address an important reliability and performance issue in TFPV modules.

To conclude, in this paper we have provided a geometrical design approach for TFPV modules, which provides a novel method for improving their shade tolerance and overall efficiency. We illustrated the geometrical aspects of partial shading, and showed how it can be overcome by breaking the symmetry in cell shape and orientation. We also demonstrate how the new cell geometries can reduce the power dissipation in the sheet conductor, using full 2D analysis for sheet resistance loss. We provide the spiral design as a realization of this design approach, which achieves the shade tolerance, as well as improved performance. We also survey the practical aspects associated with implementation of this approach, and find the state of the art instrumentation is fully capable of implementing these designs, without incurring significant losses in normal performance parameters.

**ACKNOWLEDGMENT**

This work was supported by Semiconductor Research Corporation – Energy Research Initiative (SRC-ERI), Network for Photovoltaic Technology (NPT). We would like to acknowledge the computational resources from Network for Computational Nanotechnology (NCN) at Purdue.

**APPENDIX A**

In order to simulate the behavior of a partially shaded module, we use a 1D equivalent circuit. We have shown in Ref. [14] that even for a shadow covering only a part of the cell area, the voltage developed across the shaded cell is quite uniform. Therefore, we can simplify the simulation of partially shaded module by creating a 1D equivalent circuit of $N_{series}$ series connected cells. The photocurrent of each of these cells is determined by the amount of area shaded $A_{sh}$, and the photogeneration current in the shaded area ($J_{ph,sh}$). This method is applicable for any arbitrary cell or shadow shape; therefore, for $N_{series}$ cells of a module with arbitrary shape and orientation $C_i$, and shadow of a give shape and size $S$ we need to find their intersection and shaded areas as

$$A_{sh,i} = a(C_i \cap S), \tag{A1}$$

where $a()$ denotes the area of a given shape, and the intersection determines the region of cell $C_i$ covered by the shadow $S$ (see schematic in Figure A1(a)). Now, we can calculate the current output of each cell by using photogeneration in shaded ($J_{ph,sh}$) and unshaded regions ($J_{ph}$) as

$$J_{sh,i} = J_{ph,sh} A_{sh,i} + J_{ph}(A_{cell} - A_{sh,i}). \tag{A2}$$

Here, we have assumed $J_{ph,sh} = 0.2 J_{ph}$ for a typical case, but a more appropriate value based on the local weather and dust conditions can be used for a more accurate analysis. This can then be used to create the 1D circuit for the module with $N_{series}$ solar cells with different photocurrent output (see Figure A1(b)), each of which is represented by an appropriate equivalent circuit depending on the technology under consideration (a-Si:H in this case [32]). We assume all cells have identical IV characteristics, with the photocurrent as the only varying parameter, depending on the amount of shading. We can simulate this partially shaded module, with external bypass diodes, in the string topology (Figure 2(a)) using SPICE, and obtain the operating voltage of each cell in the partially shaded module, for any given shadow dimensions. The minimum of these cell voltages ($V_{cell}^{min}$) is calculated for all possible shading configurations. This value is compared for different designs, as a measure of its shade tolerance. From this simulation, we also obtain the DC power output of the string, for different shading conditions, and can identify when external bypass will turn on to clamp the loss in power output. Note that in the circuit simulation, the series resistances connecting all cells are kept constant for all three designs. While this is not strictly the case for non-rectangular cells, it has negligible impact on shadow effects. This is because, the current flow in the sheet conductors in the radial and spiral designs is two dimensional. Therefore, a single net resistance for whole cell cannot be used, and we must use a full continuity equation solution to determine the resistive dissipation in these cell geometries, as discussed in Appendix B.

**APPENDIX B**

In order to analyze current flow in a cell of arbitrary shape, we must consider the 2D continuity equation for current in the TCO and metal layers. This can be done by solving a set of coupled continuity equations for both 2D sheet conductors, so that

$$\begin{bmatrix} \vec{\nabla}_{xy} \cdot \vec{J}_{xy}^{TCO} \\ \vec{\nabla}_{xy} \cdot \vec{J}_{xy}^{metal} \end{bmatrix} = \begin{bmatrix} J_{ph} \\ -J_{ph} \end{bmatrix}. \tag{B1}$$

Here, $\vec{\nabla}_{xy}$ is the divergence in 2D, $\vec{J}_{xy}$ is the sheet current per unit width ($A/cm$), and $J_{ph}$ is the photocurrent density in $A/cm^2$, which is being injected in plane at all points on the TCO or metal. The negative sign of $J_{ph}$ denotes current exiting the metal layer into the solar cell, and the positive sign reflects current entering the TCO layer from the solar cell. The local magnitude of the photocurrent is a function of local potential difference between the TCO and metal; i.e. $J_{ph} = f(\phi_{metal}^{TCO})$, where $\phi_{metal}^{TCO}$ is the voltage difference between the two contacts, and $f(\ )$ stands for the solar cell IV characteristics. This system of couple equations can be simplified considerably, however, if we assume the metal to be far more conductive than the TCO. Now the metal layer can be assumed equipotential and we only need to solve one continuity equation for the TCO layer, as

$$\vec{\nabla}_{xy} \cdot \vec{J}_{xy} = J_{ph}. \tag{B2}$$

A further simplification is possible, if we assume $J_{ph}$ to be constant in the voltage range of interest and set it equal to $J_{ph}$ at the maximum power point ($J_{ph}^{mpp}$) of an ideal cell ($13 mA/cm^2$ for the a-Si:H technology considered). Note that these simplifications have very little effect on the accuracy of the calculation of sheet resistance loss. Moreover, the error is same across all cell geometries, and will not affect the comparison of different cell geometries.

Now, we use the TCO sheet resistance to write $\vec{J}_{xy} = \sigma_{xy}\vec{\mathcal{E}}_{xy}$, for sheet conductor with conductivity $\sigma$ and the in-plane electric field $\vec{\mathcal{E}}_{xy}$. And, using the relation $\sigma_{xy} = 1/R_\square$ in 2D, we can write the equation in terms of contact sheet resistance as $\vec{\nabla}_{xy} \cdot \vec{\mathcal{E}}_{xy} = J_{ph}R_\square$. Finally, converting this to a voltage using $\vec{\mathcal{E}}_{xy} = -\nabla_{xy}\phi$ we can get

$$-\nabla^2_{xy}\phi = J_{ph}R_\square. \tag{B3}$$

In this setup current is injected at all points of the TCO, and exits at one of the edges, which is connected to the metal contact of the next cell. It is assumed that metal is highly conductive, and the voltage at this edge is kept constant to $V_{cell}^{mpp}$ (Dirichlet condition); while all other boundaries are at open circuit condition (Neumann condition), as shown in the schematic in Figure B1(a). Given these boundary conditions, this equation can be solved numerically, for any arbitrary 2D geometry with a finite element PDE solver, and the total power dissipation can be calculated using Equation (2). This formulation is similar to the one used for crystalline cells [33], [34], and is a generalization of the piecewise circuit approach used for rectangular solar cells [35], [36]. It can be shown easily that for the rectangular cells, the solution to Equation (B3), reproduces the results of the piecewise approach. As shown in the schematic in Figure B1, for the rectangular cell, there is no current flow in the horizontal (x) direction. Therefore, we can write the solution to Equation (B3) in 1D as

$$\phi(x,y) = -\frac{J_{ph}R_\square}{2}y^2 + k_1 y + k_2, \tag{B4}$$

where $k_1, k_2$ are the constants to be determined by the boundary conditions. Therefore, the electric field is given by $\vec{\mathcal{E}} = -\nabla_{xy}\phi = (J_{ph}R_\square y + k_1)\hat{y}$. Applying open circuit boundary condition at $y = 0$, we have $\vec{J}(x,0) = \vec{\mathcal{E}}(x,0) = 0$, which yields $k_1 = 0$. Now, we can find the power dissipation over the cell area using Equation (2) as

$$P_{dis}^{rec} = \frac{1}{R_\square} \iint\limits_{x=0,y=0}^{x=w_{cell},y=l_{cell}} J_{ph}^2 R_\square^2 y^2 dx dy = \frac{J_{ph}^2 R_\square}{3} l_{cell}^3 w_{cell} = A_{cell} \frac{J_{ph}^2 R_\square}{3} l_{cell}^2. \tag{B5}$$

This is exactly equal to the result obtained by taking a limit on the piecewise equivalent circuit approach in [35], [36]. We also use this analytical calculation to check our numerical simulation for the rectangular case, so that the simulations can be used reliably for more complicated geometries.

Figure B2(b) shows the power dissipation per unit area for the cells with rectangular, triangular and polygonal geometries respectively. From these plots, we can see that the power dissipation is dominated by the regions near the current colleting edge (connected to the metal contact). This is because, the sheet conductor near the contact carries most of the current generated within the cell area, and hence dissipates most of power. This is very apparent for the triangular cells, where the cell width is larger towards the outer edges, and the corresponding power dissipation is also higher (Figure B1(b)). This asymmetry in dissipation profiles is exploited in the spiral design to reduce overall resistive loss. As the cell shape is elongated while keeping the area constant (Figure B1(a)); the cross sectional width is reduced and each unit distance of the current collecting edge collects current from a smaller region which reduces local power dissipation (see Figure B1(b)). This reduces the overall resistive power loss in these curvilinear cells by a significant amount. This can also be seen qualitatively from Equation (B5), which has as square dependence on $l_{cell}$. In triangular geometry, where the average cross sectional distance is between 0 to $2l_{cell}$, the wider regions dissipate four times as much as thinner regions causing higher losses in total. This problem is averted for the spiral design because, the cells are longer and thinner, and the average cross sectional distance stays below $l_{cell}$, which suppresses the total dissipation.

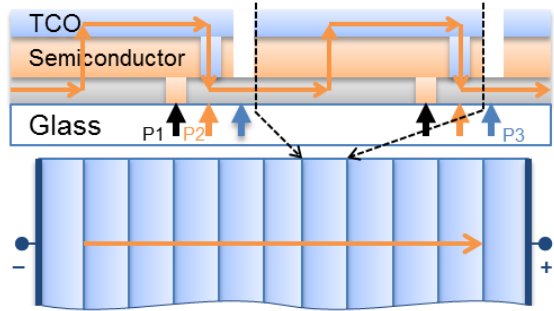

**Figure 1.** Schematics showing the side view of TFPV module, showing the laser scribes (P1/P2/P3) and various layers used for creating the series connections; and the top view of resulting series connected (arrows show direction of current flow) cells shows rectangular cells forming the module (below).

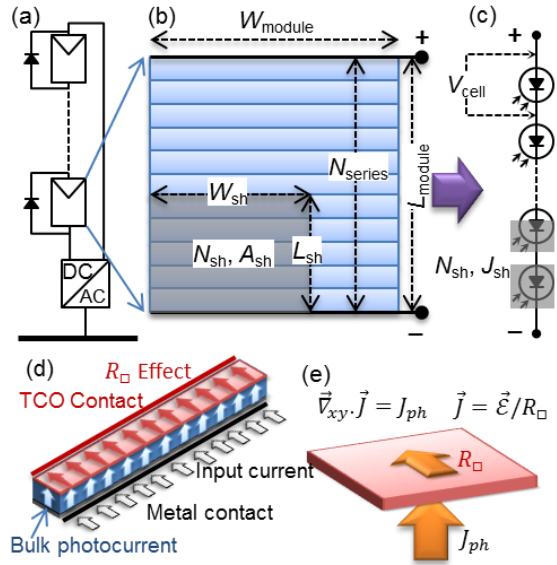

**Figure 2.** (a) String configuration considered in this study, showing modules with external bypass diodes, connected in series to a string inverter connected to system bus. (b) Schematic of a typical TFPV module with rectangular series connected cells, with a partial shadow, showing module and shadow dimensions. (c) One dimensional equivalent circuit of the partially shaded module, showing the cell voltage, and number of shaded/unshaded cells, as well as photocurrent output of shaded cells. (d) 3D schematic of an individual cell showing current conduction from the back metal contact, to bulk current and finally current collection from the edge of TCO contact. (e) Continuity and current conduction equations for a TCO top contact with current injection from the bulk solar cell beneath.

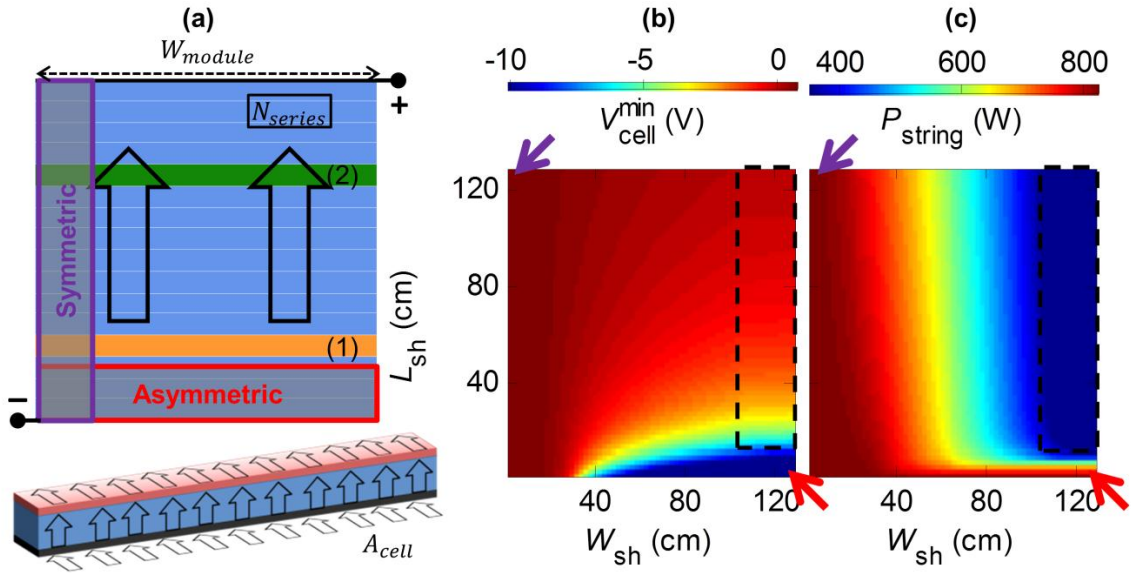

**Figure 3.** (a) Schematic of a typical TFPV module with rectangular cells. Arrows show the direction of current flow in $N_{series}$ series connected cells, each with area $A_{cell}$ (see bottom for 3D current flow patter at cell level). 2D color plots of (b) minimum cell voltage $V_{cell}^{min}$, and (c) string output power ($P_{string}$) for all possible rectilinear shadows on a typical rectangular module. Each point on the plot corresponds to a shadow of length $L_{sh}$ and width $W_{sh}$, and the color denotes the worst case reverse stress $V_{cell}^{min}$, or power output $P_{string}$ (see color bar). The dashed box highlights the cases where external bypass is on. Schematic in (a) also defines symmetric shadow (magenta), and asymmetric shading (red); the corresponding $V_{cell}^{min}$ and $P_{string}$ are highlighted with arrows on respective plots.

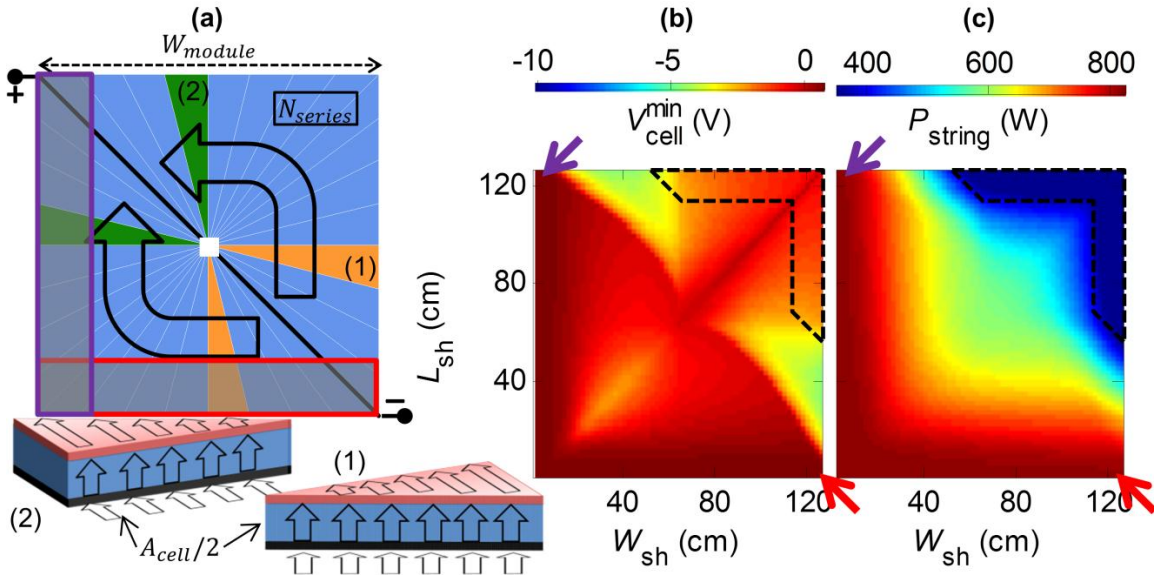

**Figure 4.** (a) Schematic showing the radial design for TFPV modules, with triangular half-cells arranged in a radial pattern, with terminals along diagonals. Each cell comprises 2 triangles with of area $A_{cell}/2$, so that the total cell area is the same (marked green e.g.). There are 2 types of triangular half-cells, depending on whether the current is collected at longer (2) or shorter (1) edge (3D current flow shown in the schematic below). (b) Color plot showing that the worst case $V_{cell}^{min}$ value for this design is restricted to $-4.7V$ (see color bar), for all possible rectilinear shadows as before. (c) The color plot of $P_{string}$ values for radial modules also show improvement for smaller shadow sizes. The cases when external bypass turns on area highlighted by the dashed polygon. The radial symmetry of design is also apparent in shade response (marked by arrows).

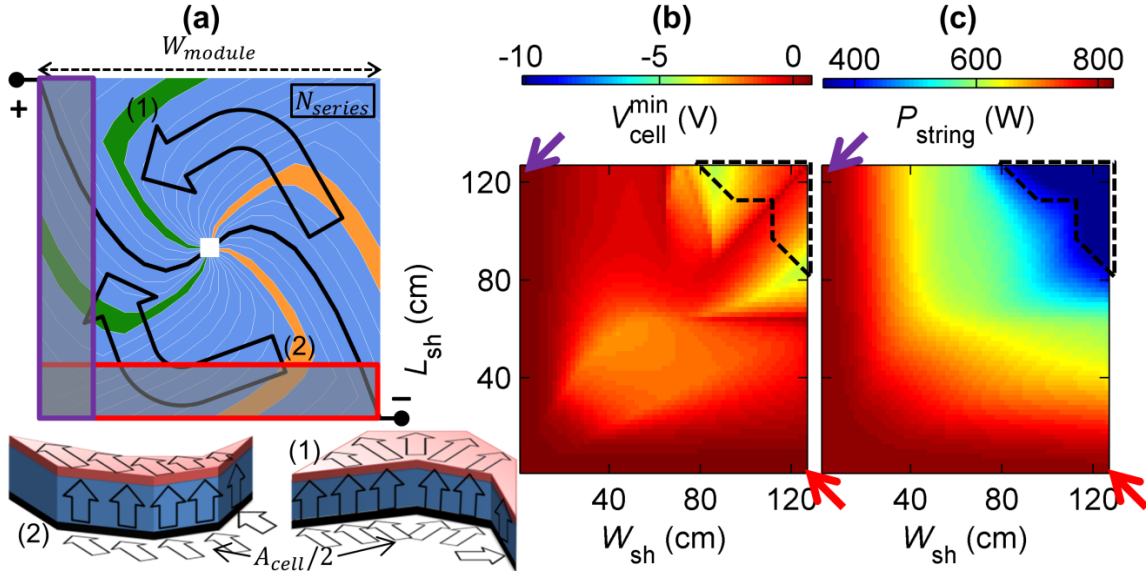

**Figure 5.** (a) Schematic showing the spiral design of TFPV module, with $N_{series}$ cells in series. Each cell consists of two half-cells (marked green e.g.) shaped like concave polygons, requiring curved terminals as shown, and resulting in the current flow direction marked by arrows. The schematic below shows 3D current flow pattern in half-cells with current flowing towards outer (1) and inner (2) edges. (b) Color plot of $V_{cell}^{min}$ values for this spiral module, for various shadow sizes, shows the worst case value to be restricted to $-4.2V$ (see color bar). (c) The color plot of string output power $P_{string}$ also shows marked improvement. The cases when external bypass turns on are highlighted by the dashed polygon. The directional symmetry of radial design is also retained (marked by arrows).

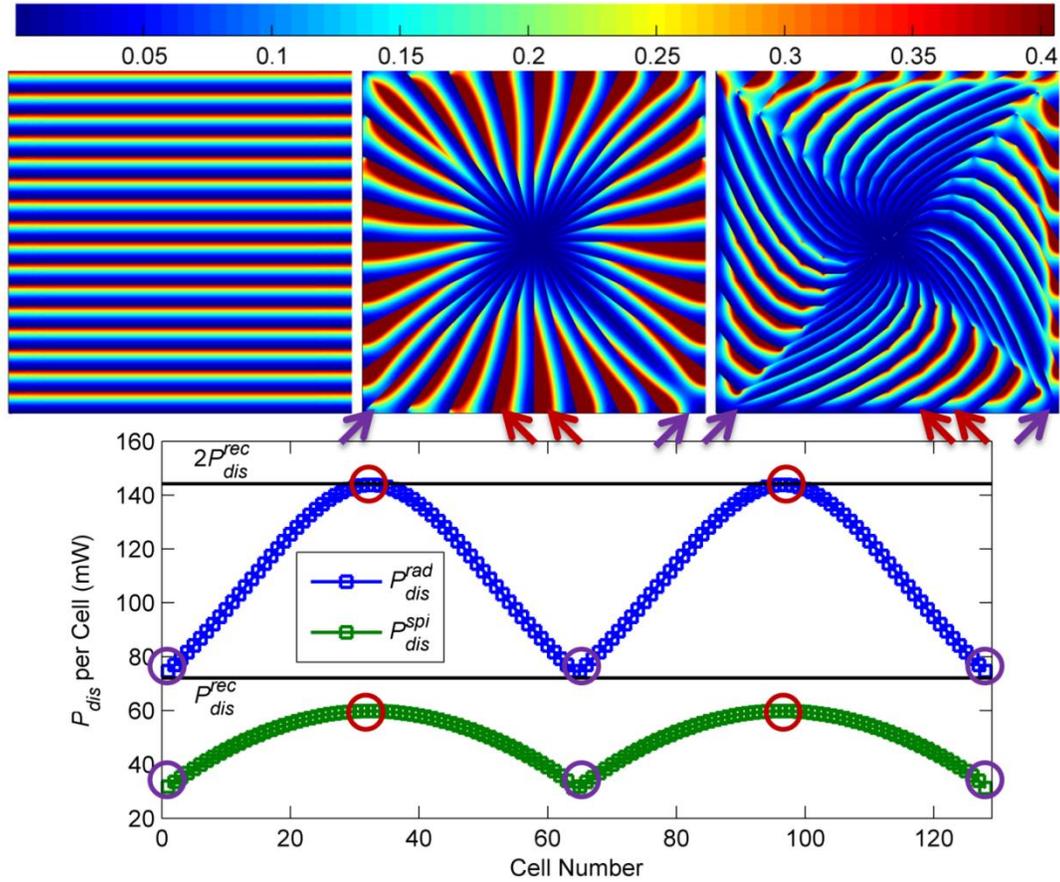

**Figure 6**. PDE simulation results for rectangular (from Figure 3(a)), radial (from Figure 4(a)), and spiral (from Figure 5(a)) sub-modules, showing distribution of resistive power dissipation per unit area in the TCO layer (color bar in $mW/cm^2$). Plot at the bottom shows power loss in each of the 128 cells for radial ($P_{dis}^{rad}$) and spiral ($P_{dis}^{spi}$) designs, compared to the dissipation in rectangular cells ($P_{dis}^{rec}$) which is same for all cells. The annotations highlight the fact that thin long cells near diagonals, have lower power dissipation (magenta); while the wider cells near the center have significantly higher power dissipation (red), for both the designs.

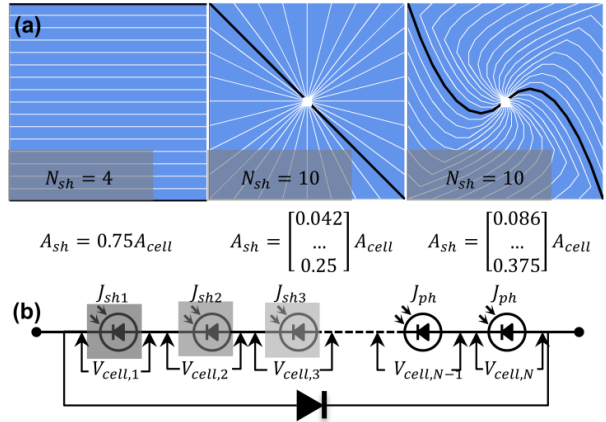

**Figure A1.** (a) Schematics showing the same size shadow on different module designs, showing the number of shaded cells ($N_{sh}$), and shaded areas $A_{sh,i}$ (identical for rectangular designs, but different for others). (b) Schematic of 1D equivalent circuit of the partially shaded module (with external bypass), using the calculated $J_{sh}$ values from the respective $A_{sh}$ values for each cell. The individual cell voltages are shown, and minimum $V_{cell}$ is calculated at each shading condition for all the designs.

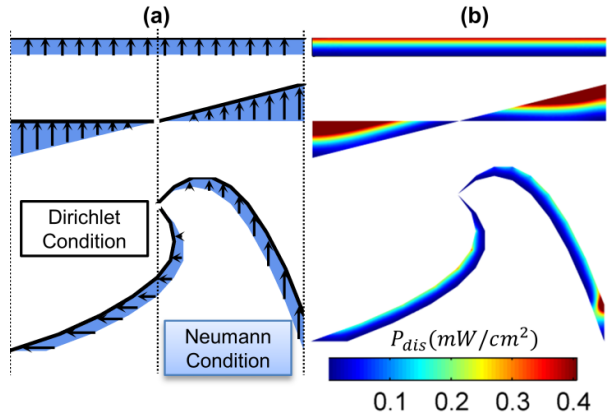

**Figure B1** (a) Schematics showing the cell shapes for the different designs, with the current collecting edge highlighted in black, and the direction of current flow shown by the arrows. (b) The distribution of resistive power dissipation per unit area for the different cells obtained from numerical solution of equation (2) (color bar in $mW/cm^2$).